\documentclass[onecolumn]{emulateapj}
\usepackage{apjfonts}

\newcommand{\be}{\begin{equation}}
\newcommand{\ee}{\end{equation}}

\def\bB{{\bf B}}
\def\eg{{\it e.g.}\ }

\def\ie{{\it i.e.}\ }
\def\vs{{\it versus}\ }

\slugcomment{\today}

\shorttitle{Multi-scale Hall-MHD turbulence in the solar wind}
\shortauthors{Galtier and Buchlin}

\begin{document}

\title{Multi-scale Hall-MHD turbulence in the solar wind}
\author{S\'ebastien Galtier}
\affil{Institut d'Astrophysique Spatiale (IAS), B\^atiment 121, 
F-91405 Orsay (France); Universit\'e Paris-Sud 11 and CNRS (UMR 8617)}
\email{sebastien.galtier@ias.fr}
\and
\author{\'Eric Buchlin}
\affil{Space and Atmospheric Physics Department, The Blackett Laboratory, 
Imperial College, London SW7 2BW, UK}

\begin{abstract}
Solar wind magnetic fluctuation spectra exhibit a significant power law steepening at frequencies
$f>1\,$Hz. The origin of this multi-scaling is investigated through dispersive Hall 
magnetohydrodynamics. We perform three-dimensional numerical simulations in the framework of 
a highly turbulent shell model and show that the large-scale magnetic fluctuations are characterized 
by a $k^{-5/3}$--type spectrum which steepens at scales smaller than the ion inertial length $d_i$, to 
$k^{-7/3}$ if the magnetic energy overtakes the kinetic energy, or to $k^{-11/3}$ in the opposite 
case. These results are in agreement both with a heuristic description {\it \`a la} Kolmogorov, and 
with the range of power law indices found in the solar wind. 
\end{abstract}


\medskip
\section{Introduction}

The interplanetary medium provides a vast natural laboratory for studying many fundamental 
questions about astrophysical plasmas. From the very beginning of {\it in situ} observations it was 
realized that this medium was not quiet but rather highly turbulent and permeated by fluctuations of 
plasma flow velocity and magnetic field on a wide range of scales, from $10^{-6}\,$Hz up to several 
hundred hertz \citep{Coleman68,Belcher71,coroniti,Bill82,Denskat83,Leamon98b,Bale}. The detailed 
analyses revealed that these fluctuations are mainly characterized (at one astronomical unit) by power 
law energy spectra around $f^{-1.7}$ at low frequency ($f<1\,$Hz), which are generally interpreted 
directly as wavenumber spectra by using the Taylor ``frozen-in flow'' hypothesis \citep{Goldstein99}. 
This spectral index is somewhat closer to the Kolmogorov prediction for neutral fluids ($-5/3$) than the 
Iroshnikov--Kraichnan prediction for magnetohydrodynamic (MHD) ($-3/2$) \citep{K41,iro,kr65}. Both 
heuristic predictions are built, in particular, on the isotropic turbulence hypothesis which is questionable 
for the inner interplanetary medium \citep{dobro,Galtier05,Oughton} since apparent signatures of 
anisotropy are found through, for example, the detection of Alfv\'en waves \citep{Belcher71} or the 
variance analysis of the magnetic field components and magnitude \citep{Barnes}. Note that from 
single-point spacecraft measurements it is clearly not possible to specify the exact three-dimensional 
(3D) nature of the interplanetary turbulent flow which still remains an open question. 

For timescales shorter than few seconds ($f>1\,$Hz), the statistical properties of the solar wind 
change drastically with, in particular, a steepening of the magnetic fluctuation power law spectra 
over more than two decades \citep{coroniti,Denskat83,Leamon98b,Bale,Smith06} with a spectral 
index on average around $-3$. 
The range of values found is significantly broader than the large scale counterpart and may depend 
on the presence of magnetic clouds which lead to less steep power laws than open magnetic field 
line regions \cite{Smith06}. 
This new inertial range -- often called dissipation range -- is characterized 
by a bias of the polarization suggesting that these fluctuations are likely to be right-hand polarized 
\citep{Golstein94} with a proton cyclotron damping of Alfv\'en left circularly polarized fluctuations 
\citep{Stawicki01}. This proposed scenario seems to be supported by Direct Numerical Simulations 
(DNS) of compressible $2{1 \over 2}$D Hall-MHD turbulence \citep{Ghosh96} where a steepening 
of the spectra is found -- although on a narrow range of wavenumbers -- and associated with 
the appearance of right circularly polarized fluctuations. It is likely that what has been conventionally 
thought of as a dissipation range is actually a second -- dispersive -- inertial range and that the steeper 
power law is due to nonlinear wave processes rather than pure dissipation \citep{krishan04}. 

In this paper, our main goal is to investigate numerically the origin of the steepening of the magnetic 
fluctuation power law spectra observed in the solar wind. For that purpose, we develop a numerical 
cascade model based on dispersive Hall-MHD. We present the model in Section \ref{sec2} and the 
numerical results in Section \ref{sec3}. A discussion about the duality between nonlinear cascade and 
kinetic dissipation is given in Section \ref{sec4}. A conclusion follows in the last section.

\medskip
\section{Hall MHD equations and cascade model}
\label{sec2}

Spacecraft measurements made in the interplanetary medium suggest a nonlinear dispersive 
mechanism that will be modeled by the 3D incompressible Hall-MHD equations. 
Such a description is often used, for example, to understand the main impact of the Hall term in 
turbulent dynamo, in the solar wind and wave turbulence \citep{krishan04,Mininni05,Galtier06}. 
It is particularly relevant for the pure/polar wind where density fluctuations are weak. 
The incompressible inviscid Hall-MHD equations read 
\be
\nabla \cdot {\bf V} = 0 \, ,\qquad
\nabla \cdot \bB = 0 \, ,
\label{hmhd1}
\ee
\be
\frac{\partial {\bf V}}{\partial t} + {\bf V} \cdot \nabla \, {\bf V} = 
- {\bf \nabla} P_* + \bB \cdot \nabla \, \bB \, ,
\label{hmhd2}
\ee
\be
\frac{\partial \bB}{\partial t} + {\bf V} \cdot \nabla \, \bB = 
\bB \cdot \nabla \, {\bf V} - d_i \, \nabla \times [ (\nabla \times \bB) \times \bB ] \, ,
\label{hmhd3}
\ee
where $\bB$ has been normalized to a velocity ($\bB \to \sqrt{\mu_0 n m_i} \, \bB$, with 
$m_i$ the ion mass and $n$ the electron density), ${\bf V}$ is the plasma flow velocity, 
$P_*$ is the total (magnetic plus kinetic) pressure and $d_i$ is the ion inertial length 
($d_i = c / \omega_{pi}$, where $c$ is the speed of light and $\omega_{pi}$ is the ion plasma 
frequency). The Hall effect appears in the induction equation as an additional term proportional 
to the ion inertial length $d_i$ which means that it is effective when the dynamical scale is small 
enough \citep{Bhatt04}. In other words, for large scale phenomena this term is negligible and 
we recover the standard MHD equations. In an opposite limit, \eg for very fast time scales 
($\ll \omega_{ci}^{-1}$, the ion cyclotron period), ions do not have time to follow electrons and 
provide a static homogeneous background on which electrons move. Such a model where 
the dynamics is entirely governed by electrons is called electron MHD \citep{Kingsep}. It can be 
recovered from Hall-MHD by taking the limits of small velocity ${\bf V}$ and large $d_i$. 

DNS of turbulent flows at very large (magnetic) Reynolds numbers are well beyond today's computing 
resources. Therefore, any reasonable simplification of corresponding equations is 
particularly attractive. In the case of the solar wind, for which the Reynolds number is as large 
as $10^9$ \cite{Tajima}, simplified models are currently the only way to investigate the multi-scale 
behavior described above. Following this idea, we propose a description of solar wind turbulence 
in terms of a shell model based on the 3D incompressible Hall-MHD equations. The basic idea of this 
shell model is to represent each spectral range of a turbulent velocity and magnetic field with a few 
variables and to describe their evolution in terms of relatively simple Ordinary Differential Equations 
(ODE), ignoring details of its spatial distribution. The form of the ODE is of course inspired from the 
original partial derivative equations and depends on some coefficients which are fixed by 
imposing the conservation of the inviscid invariants. In spite of the simplifications made, 
shell models remain highly non trivial and are able to reproduce several aspects of turbulent flows 
like intermittency \citep{frisch,Bife,Buchlin06}. 
Shell models are however less relevant in situations where strong non local interactions dominate and, 
of course, when information in the physical space is necessary. Anisotropy is also problem for cascade 
models such the one used in this paper, nevertheless it may be described by shell models if they are 
derived, for example, from spectral closure like EDQNM \cite{Carbone90}. 

The present shell model is governed by the following coupled nonlinear ODE equations \citep{hori}
\be
\label{shell1}
\frac{\partial V_n}{\partial t} + \nu_2 k_n^4 V_n= 
i k_n \left[ ( V_{n+1} V_{n+2} - B_{n+1} B_{n+2} ) - \frac{1}{4} ( V_{n-1} V_{n+1} - 
B_{n-1} B_{n+1} ) - \frac{1}{8} ( V_{n-2} V_{n-1} - B_{n-2} B_{n-1} ) \right]^* \, ,
\ee
\be
\label{shell2}
\frac{\partial B_n}{\partial t} + \eta_2 k_n^4 B_n =
\frac{i k_n}{6} \left[ ( V_{n+1} B_{n+2} - B_{n+1} V_{n+2} ) + ( V_{n-1} B_{n+1} - 
B_{n-1} V_{n+1} ) + ( V_{n-2} B_{n-1} - B_{n-2} V_{n-1} ) \right]^*
\ee
$$+ (-1)^n i d_i k_n^2 \left[ B_{n+1} B_{n+2} - \frac{1}{4} B_{n-1} B_{n+1} - 
\frac{1}{8} B_{n-2} B_{n-1} \right]^* \, ,
$$
where $^*$ stands for the complex conjugate. The complex variables $V_n(t)$ and $B_n(t)$
represent the time evolution of the field fluctuations over a wavelength $k_n = k_0 \lambda^n$, with 
$\lambda \equiv 2$ the intershell ratio and $n$ varying between $1$ and $N$. We note immediately 
that the present model tends to the well-known shell model for MHD \citep{Frick,Giuliani} when the 
large scale limit is taken, \ie in the limit $k_n d_i \to 0$. Note also the use of hyperviscosities 
($\nu_2$, $\eta_2$) to extend at maximum the nonlinear dispersive inertial range. The dissipation 
is mainly used for numerical stability since the solar wind is mainly collisionless. We focus our 
attention only on the wavenumber scales where dissipation is negligible therefore we do not 
investigate the exact form of the dissipation. 

By construction, equations (\ref{shell1}--\ref{shell2}) conserve the three inviscid invariants of incompressible 
Hall-MHD \citep[see \eg][]{Galtier06}
\be
E = \int E(k) \, dk = \frac{1}{2} \sum_n ( \vert V_n \vert^2 + \vert B_n \vert^2 ) = \sum_n E(k_n) \, ,
\label{energy}
\ee
\be
H_m = \int H_m(k) \, dk = \frac{1}{2}  \sum_n (-1)^n \frac{\vert B_n \vert^2}{k_n}  = \sum_n H_m(k_n) \, ,
\ee
\be
H_h = \int H_h(k) \, dk = \frac{1}{2}  \sum_n \left[ (-1)^n  d_i^2 k_n \vert V_n \vert^2 + 
d_i (V_n^* B_n + V_n B_n^* ) \right] =\sum_n H_h(k_n) \, ,
\ee
which are respectively the total energy, the magnetic and hybrid helicities. Note, as usual, a 
difference of unity in wavenumber between the shell (in $k_n$) and the true (in $k$) power 
spectra \citep{Frick,Giuliani}.

\clearpage

From equations (\ref{shell1},\ref{shell2},\ref{energy}) it is possible 
to extract information about the energy flux $P_n$ towards small scales
\cite{Buchlin06}.
We have (for an infinite range of shell indices) 
\begin{eqnarray}
P_n &=& - \frac12 \sum_{m \ge n} \Bigg\{ i \frac{k_m}{8} 
\Big[8(V_mV_{m+1}V_{m+2} - V_mB_{m+1}B_{m+2} ) 
- 2(V_{m-1}V_{m}V_{m+1} - B_{m-1}V_{m}B_{m+1} )
- (V_{m-2}V_{m-1}V_{m} - B_{m-2}B_{m-1}V_{m} )\Big] \nonumber \\
&&\quad+ \frac{i k_m}{6} \Big[ (B_{m}V_{m+1}B_{m+2} - B_{m}B_{m+1}V_{m+2}) 
+ (V_{m-1}B_{m}B_{m+1} - B_{m-1}B_{m}V_{m+1})  
+ (V_{m-2}B_{m-1}B_{m} - B_{m-2}V_{m-1}B_{m}) \Big] \nonumber \\
&&\quad + (-1)^m i d_i \frac{k_m^2}{8}
\Big[8B_mB_{m+1}B_{m+2} - 2B_{m-1}B_{m}B_{m+1} - B_{m-2}B_{m-1}B_{m}
\Big] \Bigg\} + c.c.  \, .
\end{eqnarray}
Simple manipulations (with $\lambda \equiv 2$) lead to 
\begin{eqnarray}
P_n &=& 
- \frac12 \Bigg\{ i \frac{k_n}{8}
\Big[- 2(V_{n-1}V_{n}V_{n+1} - B_{n-1}V_{n}B_{n+1} )
- (V_{n-2}V_{n-1}V_{n} - B_{n-2}B_{n-1}V_{n} ) 
- 2(V_{n-1}V_{n}V_{n+1} - B_{n-1}B_{n}V_{n+1} )\Big] \nonumber \\
&&+ \frac{i k_n}{6} \Big[ (V_{n-1}B_{n}B_{n+1} - B_{n-1}B_{n}V_{n+1}) 
+ (V_{n-2}B_{n-1}B_{n} - B_{n-2}V_{n-1}B_{n}) + 2 (V_{n-1}B_{n}B_{n+1} - 
B_{n-1}V_{n}B_{n+1}) \Big] \nonumber \\
&&\quad + (-1)^n i d_i \frac{k_n^2}{8}
\Big[ -2 B_{n-1}B_{n}B_{n+1} - B_{n-2}B_{n-1}B_{n} + 
4 B_{n-1}B_{n}B_{n+1}\Big] \Bigg\} \nonumber \\
&&- \frac12 \sum_{m \ge n}
\Bigg\{ i \frac{k_m}{2} \Big[2(V_mV_{m+1}V_{m+2} - V_mB_{m+1}B_{m+2} ) 
- (V_{m}V_{m+1}V_{m+2} - B_{m}V_{m+1}B_{m+2} ) - (V_{m}V_{m+1}V_{m+2} - B_{m}B_{m+1}V_{m+2} )\Big] \nonumber \\
&&\quad+ \frac{i k_m}{6} \Big[ (B_{m}V_{m+1}B_{m+2} - B_{m}B_{m+1}V_{m+2}) 
+ 2 (V_{m}B_{m+1}B_{m+2} - B_{m}B_{m+1}V_{m+2}) + 4 (V_{m}B_{m+1}B_{m+2} - B_{m}V_{m+1}B_{m+2}) \Big] \nonumber \\
&&\quad + (-1)^m i d_i k_m^2
\Big[B_mB_{m+1}B_{m+2} + B_{m}B_{m+1}B_{m+2} - 2 B_{m}B_{m+1}B_{m+2}\Big]  \Bigg\}
+ c.c. \, .
\end{eqnarray}
All terms in the sum over $m$ vanish and we finally obtain after rearranging the remaining terms
\begin{eqnarray}
P_n &=& 
\label{flux}
- i \frac{k_n}{48} \Big[ 4 V_{n-1}B_{n} ( 3B_{n+1} - B_{n-2} )
- 3 V_{n-1}V_{n} ( V_{n-2} + 4 V_{n+1} ) 
+ 2 B_{n-1} B_{n} ( V_{n+1} + 2 V_{n-2} ) 
- B_{n-1} V_{n} ( 2 B_{n+1} - 3 B_{n-2} ) \Big]  \nonumber \\
&&- i d_i \frac{k_n^2}{16} (-1)^n B_{n-1}B_{n} ( 2 B_{n+1} - B_{n-2} )
+ c.c. \, .
\end{eqnarray}
This previous expression allows us to derive one particular type of solutions of constant flux 
towards small scales by assuming that the velocity and magnetic fields are power law 
dependent in $k_n$, namely
\be
V_n \sim k_n^{\alpha} \sim \lambda^{n \alpha} \, ,
\qquad
B_n \sim k_n^{\beta} \sim \lambda^{n \beta} \, . 
\ee
We insert the previous relations into equation (\ref{flux}) and cancel the dependence in $n$ 
to obtain a constant flux solution. It gives 
\be
3 \alpha +1 =0 \, ,
\label{relation1}
\ee
\be
1 + \alpha + 2 \beta =0 \, ,
\qquad
3 \beta + 2 =0 \, . 
\ee
The last relation which comes from the Hall term leads to a $k^{-7/3}$ magnetic energy spectrum. 
It is precisely the expected scaling exponent in electron MHD turbulence when the magnetic field dominates 
at small scales \citep{Biskamp96}. The two other relations, applicable in particular at large scales in the pure 
MHD turbulence regime, lead to a unique $k^{-5/3}$ scaling for magnetic and 
kinetic energy spectra. Note that equation (\ref{relation1}) comes from the pure velocity interacting 
term of equation (\ref{shell1}): in other words it is the Navier-Stokes contribution to Hall-MHD which, 
as expected, gives the Kolmogorov scaling exponent. 

\section{Numerical results}
\label{sec3}

Generally, shell models do not deal with spectral anisotropy, 
therefore we will focus our analysis on the isotropic spectral evolution when the Hall term is effective. 
To our knowledge, such an analysis has never been made with a shell model. As explained above 
some DNS exist but the resolution is currently limited at maximum to a spatial resolution of $256^3$ 
grid points \citep{Ghosh96,Mininni05} which is already interesting to analyze first dispersive effects 
but definitely not enough to extract precisely any multi-scale spectral power law behaviors. 

\begin{figure}[t]
\centering
\includegraphics[width=.55\linewidth]{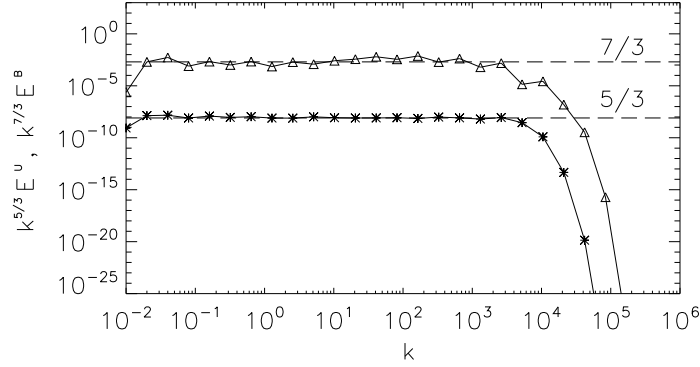}
\caption{Compensated magnetic (triangles) and kinetic (stars) energy spectra for, respectively, 
the electron MHD and Navier-Stokes limits. The corresponding well-known scalings in $k^{-7/3}$ 
and in $k^{-5/3}$ are given in dashed lines.}
\label{Fig1}
\end{figure}

\subsection{Electron MHD and Navier-Stokes limits}

Numerical simulations of equations (\ref{shell1}--\ref{shell2}) are made with $N=25$, $k_0=10^{-2}$ and 
without external forcing. In all cases considered in this paper, the initial spectra are localized at large 
scales with a maximum around $k=0.04$ and with an sharp decrease at larger wavenumbers. First, we 
consider the purely magnetic case also called electron MHD ($V_n=0$ at any time, $d_i = 0.3$ and 
$\eta_2=10^{-13}$). The compensated magnetic energy spectrum is shown in Fig. \ref{Fig1}. 
(A time average is taken over 30 times in all figures.) 
As expected, the magnetic energy spectrum (triangles) scales in $k^{-7/3}$ which is the Kolmogorov 
scaling counterpart for electron MHD \citep{Biskamp96}. This result differs clearly from the 
purely hydrodynamic case ($B_n=0$ at any time; $\nu_2=10^{-13}$), the stars in the same figure, 
for which we have a $k^{-5/3}$ power law. Note that MHD simulations with $d_i=0$ (not shown here) 
reproduce correctly the $k^{-5/3}$ energy spectra \citep{Frick,Giuliani}. Note also that in both cases 
(and for all other figures) the true spectra (in $k$) are displayed. From these first results, we may 
conclude naively that in Hall-MHD the magnetic energy spectrum should lie between these two scalings. 
We will see that, in general, it is not true. 

\begin{figure}[b]
\centering
\includegraphics[width=.55\linewidth]{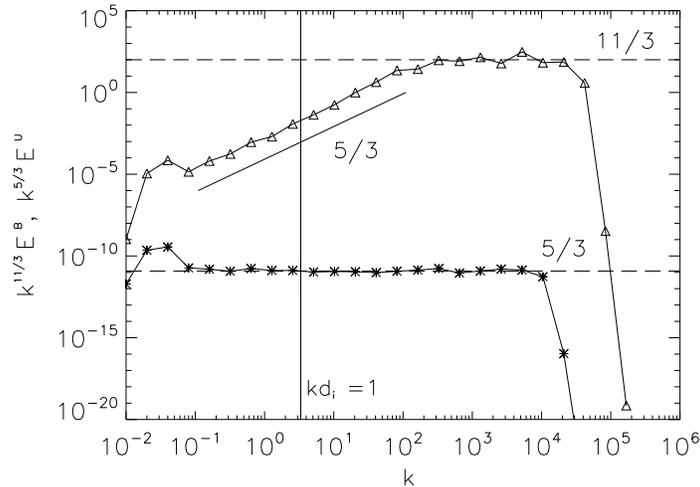}
\caption{Compensated magnetic (triangles) and kinetic (stars; for clarity, they are shifted to lower values) 
energy spectra in Hall-MHD. The vertical solid lines indicate the critical value $k d_i = 1$.}
\label{Fig2}
\end{figure}

\subsection{Hall-MHD with $d_i=0.3$}
\begin{figure}[t]
\centering
\includegraphics[width=.55\linewidth]{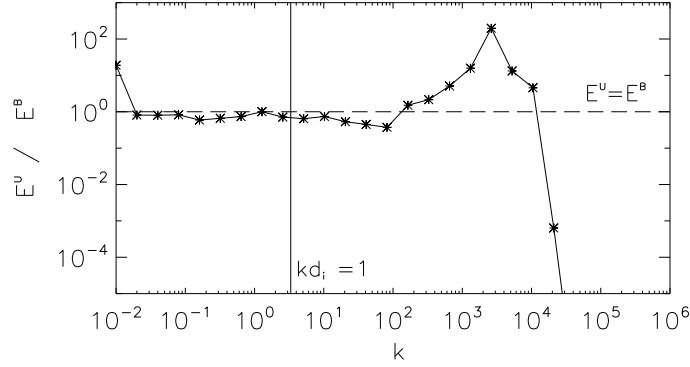}
\caption{Spectral ratio between the kinetic and magnetic energies (same simulation as in Fig. \ref{Fig2}). 
The equipartition state is given in dashed line.}
\label{Fig3}
\end{figure}

Then, we perform a full Hall-MHD numerical simulation in which the kinetic and magnetic fluctuations 
are initially of order one ($\nu_2=\eta_2=10^{-13}$ and $d_i = 0.3$). In Fig. \ref{Fig2} we show the 
magnetic and kinetic compensated energy spectra. Two scalings are clearly present 
for the magnetic energy spectrum: large scales are characterized by a Kolmogorov type spectrum 
in $k^{-5/3}$ and, surprisingly, small scales follow a $k^{-11/3}$ power law over more than two decades. 
This second inertial range appears only when $k d_i > 20$: in other words, that means the Hall term 
becomes dominant not immediately beyond the critical value $k d_i = 1$ but at scales one order of 
magnitude smaller. Note this additional difficulty for DNS to reproduce such a behavior since it requires 
to have a very extended inertial range. The kinetic part seems not to be affected by the Hall term and 
displays clearly a $k^{-5/3}$ scaling all over the wavenumbers. 
As we see 
in Fig. \ref{Fig3}, this behavior is linked to the spectral ratio between the kinetic and magnetic energies. 

The magnetic energy is slightly greater than the kinetic one at large scales, as usually found in MHD 
DNS \citep{Politano} and in the solar wind \citep{Bavassano}. This feature extends beyond the critical 
value $k d_i = 1$. Then, the kinetic energy dominates strongly the magnetic one until the dissipative 
range is reached ($k>10^4$). This result reveals that the small scale nonlinear dynamics is likely to 
be dominated by the velocity and not by the magnetic field as it is the case in the electron MHD regime. 
Finally, we have also computed the residual energy spectrum (not shown), \ie the difference in 
absolute value between the magnetic and kinetic energy spectra. This quantity follows a $k^{-5/3}$ 
power law which is clearly different from the $k^{-7/3}$ scaling found recently in pure 3D MHD DNS 
\cite{Muller}. The Hall term could be at the origin of this difference: for example, in the context of 
wave turbulence we know \cite{Galtier06} that the equipartition found in pure MHD is not anymore 
observed when the Hall term is present, whatever its magnitude is, leading to a non trivial interaction 
between the magnetic and kinetic energy spectra. 

To explain the non-trivial behavior found above, we have to come back to the original Hall-MHD equations 
(\ref{hmhd1}--\ref{hmhd3}). At large scales, the usual Kolmogorov phenomenology may be used to 
describe turbulence. We will not enter in the debate about the Kolmogorov \vs Iroshnikov-Kraichnan 
description since our primary aim is to look at the multi-scale behavior of the Hall-MHD flow and not 
the very precise value of the power law exponent at large scales. The kinetic and magnetic energies 
being of the same order of magnitude, we find from equations (\ref{hmhd2}--\ref{hmhd3}) a single 
transfer time $\tau_{tr}=\ell / V_{\ell}$ and therefore a $k^{-5/3}$ large scale energy spectrum 
\citep{frisch}. We note immediately that at small scales this time will not change for equation 
(\ref{hmhd2}) since then the velocity field dominates. However, for equation (\ref{hmhd3}) the Hall 
term has to be taken into account when scales are smaller than the ion inertial length; it gives 
\be
\tau_{tr}=\ell^2 / (d_i B_{\ell}) \, .
\ee
Equating both transfer times we obtain the relation 
\be
d_i B_{\ell} = \ell V_{\ell} \, .
\ee
Because at small scales the velocity field overtakes the magnetic field, the latter is driven nonlinearly 
by the former which eventually leads to the relation 
\be
E^B(k) = (d_i k)^{-2} E^V(k) \sim k^{-11/3} \, .
\ee
As we have seen above this result cannot be predicted by a simple analysis on constant flux solutions.

\begin{figure}[b]
\centering
\includegraphics[width=.55\linewidth]{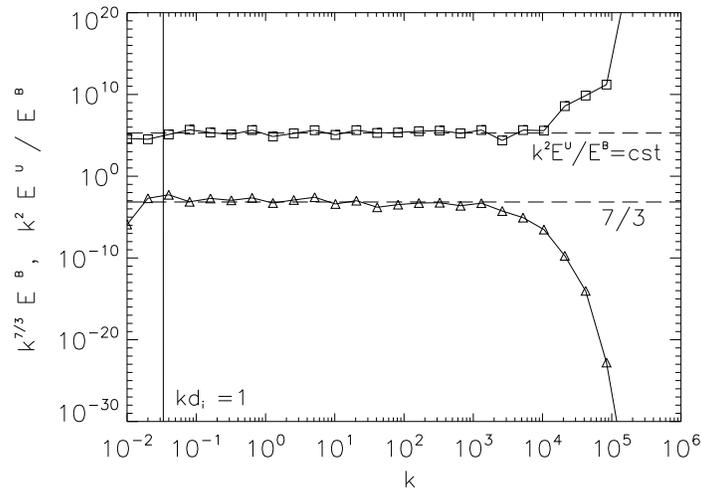}
\caption{Magnetic energy spectrum (triangles) and compensated spectral ratio (squares; for clarity, 
they are shifted to higher values) for $d_i = 30$.}
\label{Fig4}
\end{figure}

\subsection{Hall-MHD with $d_i=30$}

The heuristic description given above may be modified for other physical conditions. In a last set of 
simulations, we have taken $d_i=30$ such that the Hall term becomes effective at the very beginning 
of the inertial range ($\nu_2=\eta_2=10^{-12}$). In this case, we see in Fig. \ref{Fig4} that the 
magnetic energy exhibits the electron MHD law in $k^{-7/3}$ whereas a spectral relation $k^2 E^U = E^B$ 
is clearly established. This result means that a steeper spectrum in $k^{-13/3}$ is found for the kinetic 
energy and that the magnetic energy overtakes the kinetic one at all scales. These results may be 
explained by modifying the previous phenomenology. Since now the magnetic field overtakes the 
velocity, the relevant transfer time in equation (\ref{hmhd3}) is given by the Hall term at all scales 
which leads to $\tau_{tr}=\ell^2 / (d_i B_{\ell})$. For equation (\ref{hmhd2}), we retain the magnetic 
nonlinear term and obtain $\tau_{tr}=\ell V_{\ell} / (B_{\ell}^2)$. Equating both times we find the relation
\be
d_i V_{\ell} \sim \ell B_{\ell}
\ee
and then 
\be
E^V(k) = (d_i k)^{-2} E^B(k) \sim k^{-13/3} \, .
\ee 
This situation may be relevant for the solar wind if a strong small scale depletion of kinetic energy is produced 
for example by proton cyclotron damping \citep{Hollweg}. 
In the context of Hall MHD turbulence, we see that a significant range of spectral indices are allowed for 
the magnetic fluctuation spectrum. This range of values, between $-7/3$ and $-11/3$, has to be compared 
with the most recent works \cite{Smith06} where the value $-2.61 \pm 0.96$ is reported for open magnetic 
field line regions.

\section{Nonlinear cascade {\it versus} kinetic dissipation}
\label{sec4}

The mechanism by which heat is deposited in the low and extended solar corona is a recurring theme 
of research in solar physics. In the solar wind case, heating perpendicular to the magnetic field is clearly 
observed for protons. This is often taken to be a signature of cyclotron damping of the turbulent 
fluctuations \citep{Hollweg,Cranmer}. The fluid and kinetic descriptions are often seen as two competing 
mechanisms and it is only during the last years that attempts have been made to reconcile both 
descriptions. One of the main problem is to quantify the balance between a nonlinear cascade, from 
large scales to small (non-MHD) scales, and cyclotron damping, which may occur at small scales. A 
ratio of order one has been proposed to explain why complete 
cyclotron absorption, and the corresponding pure magnetic helicity signature, is usually not observed 
\citep{Leamon98a}. In view of the weak density fluctuations and the low average turbulent Mach 
number \citep{Bill90}, this type of analysis is generally made with a leading order description based 
on incompressible turbulence like in the present paper. 
The role of anisotropy has also been recently discussed \citep{Leamon00}: it is proposed that 
a significant fraction of dissipation likely proceeds through a perpendicular cascade and small-scale 
reconnection. The scale at which dissipation occurs is associated to the ion inertial length $d_i$ which is 
of the order of $100$km at $1$ AU. 
In the mean time, indirect mechanism for damping quasi 2D fluctuations have been proposed to explain 
the steepening of the magnetic fluctuation spectra \citep{Marko}. Indeed, whereas quasi 2D fluctuations 
dominate strongly the slab component in the MHD inertial range, it is more balanced in the 
dispersive range which suggests that most of the energy dissipated comes from the quasi 2D fluctuations. 
In that context, an equation for the energy transfer in the solar wind is proposed in an {\it ad-hoc} way where 
the diffusion and dissipative coefficients are chosen initially to produce the expected scaling laws. This 
philosophy is clearly different to the one followed here with the cascade model. 

Recently, a rigorous analysis of nonlinear transfers in the inner solar wind has been proposed in the context 
of Hall-MHD wave turbulence \citep{Galtier06}. This approach reconciles somehow the picture, in one hand, 
of a solar wind made of propagating Alfv\'en waves and, in other hand, a fully turbulent interplanetary medium.
The main rigorous result is a steepening of the anisotropic magnetic fluctuation spectrum at scales smaller 
than $d_i$ with anisotropies of different strength, large scale anisotropy being stronger than at small 
scales. This result is particularly interesting for cyclotron damping since this mechanism is thought to be 
less efficient when spectral anisotropy is stronger. Of course, Hall-MHD does not deal with kinetic effects 
and it is only a way to quantify nonlinear transfers. In the present work, we have seen that the steepening 
of the magnetic fluctuation power law spectra may be explained by a pure nonlinear mechanism. Different 
values are found for the power
law exponent which depends on the ratio between the kinetic and magnetic energies. In the previous works 
quoted above, a balance is often assumed between the kinetic and magnetic energies. This assumption is 
not necessarily satisfied and the range of values of the power exponent recently investigated \citep{Smith06} 
may be seen as a signature of different ratio between the kinetic and magnetic energies. Note that in the 
context of Hall MHD turbulence, it is straightforward to show with a heuristic model that the cascade rate 
should increase at small scales because of the Hall term. This prediction compares favorably with solar wind 
analysis showing that a steeper spectrum results from greater cascade rates \citep{Smith06}.

\section{Discussion and conclusion}
\label{sec5}

Hall-MHD may be seen as a natural nonlinear model for explaining the strong steepening of the magnetic fluctuation spectra observed in the solar wind and the precise value of this power law exponent appears 
to be a way to probe the velocity scaling law. In particular, our analysis reveals that (i) the presence of a 
wide MHD inertial range has a deep impact on smaller dispersive scales in fixing the corresponding 
spectral scaling laws, (ii) the electron MHD approximation may not be relevant for describing small scale 
solar wind turbulence, and (iii) the non trivial multi-scaling found may be seen as a consequence of the 
propagation of some large scale information to smaller scales. 

Of course, the scalings found here may be altered by effects not included in the model. For example 
density variations -- although weak in the pure/polar wind -- could modify slightly these results 
like in MHD \citep{Zank1,Zank2}, 
as well as intermittency whose effects is mainly measured in higher order moments. Nonlocal effects 
\citep{Mininni06} and anisotropy \citep{Galtier06} are also important ingredients, however in the 
latter case a recent analysis made with Cluster, a multi-spacecraft mission dedicated to the Earth's 
magnetosphere, shows only a slight difference in the power law index between the frequency magnetic 
spectrum and the 3D spatial one although a strong anisotropy is detected in this medium \citep{Sahraoui}. 
The predominance of outward propagating Alfv\'en {\it and} whistler waves has also certainly an influence 
on the spectral laws but it has never been studied in a multi-scale model and it is currently under investigation. 
It is likely that such asymmetric wave flux (imbalanced turbulence) leads for the scaling exponents to a 
range of values centered around the exponents found here as it is observed in MHD turbulence
\citep[see \eg][]{Politano, Galtier00}. In that sense the present work lays the foundation to a more 
general multi-scale model.

\acknowledgments

Grant from PPARC and partial financial support from the PNST (Programme National Soleil--Terre) 
program of INSU (CNRS) are gratefully acknowledged.

\end{document}